\documentclass[showpacs,amsmath,amssymb,twocolumn,prl,superscriptaddress]{revtex4-2}
\usepackage{amssymb}
\usepackage[dvips]{graphicx}
\usepackage{enumerate}
\usepackage{epsfig} 
\usepackage{subfigure}
\usepackage{xcolor}
\usepackage[T1]{fontenc}
\usepackage{fullpage}
\usepackage{amsthm,amsfonts,amssymb,amscd,mathrsfs,xspace,framed}
\usepackage{mathrsfs,amsmath}
\usepackage{color}
\usepackage{setspace}
\usepackage{url}
\usepackage{wrapfig}
\usepackage{tikz}
\usepackage{enumitem}
\usepackage{bm}
\usepackage{comment}
\usepackage{txfonts}
\usepackage{array}
\usepackage{color}
\usepackage{braket}
\usepackage{hyperref}
\usepackage[english]{babel}
\usepackage{url}
\usepackage{graphicx}
\usepackage{pdfpages}

\newcommand{\ZZ}[1]{{{\textcolor{black}{#1}}}}
\def\be{\begin{equation}}
\def\ee{\end{equation}}
\def\ba{\begin{eqnarray}}
\def\ea{\end{eqnarray}}

\makeatletter
\AtBeginDocument{\let\LS@rot\@undefined}
\makeatother

\def\be{\begin{equation}}
\def\ee{\end{equation}}
\def\ba{\begin{eqnarray}}
\def\ea{\end{eqnarray}}

\begin{document}

\title{Entanglement-Assisted Communication Surpassing the Ultimate Classical Capacity}

\author{Shuhong Hao}
\affiliation{
Department of Materials Science and Engineering, University of Arizona, Tucson, Arizona 85721, USA
}

\author{Haowei Shi}
\affiliation{
James C. Wyant College of Optical Sciences, University of Arizona, Tucson, Arizona 85721, USA
}

\author{Wei Li}
\affiliation{
Department of Materials Science and Engineering, University of Arizona, Tucson, Arizona 85721, USA
}

\author{Quntao Zhuang}
\affiliation{
Department of Electrical and Computer Engineering, University of Arizona, Tucson, Arizona 85721, USA
}
\affiliation{
James C. Wyant College of Optical Sciences, University of Arizona, Tucson, Arizona 85721, USA
}

\author{Zheshen Zhang}
\email{zsz@arizona.edu}
\affiliation{
Department of Materials Science and Engineering, University of Arizona, Tucson, Arizona 85721, USA
}
\affiliation{
James C. Wyant College of Optical Sciences, University of Arizona, Tucson, Arizona 85721, USA
}

\begin{abstract}
Entanglement underpins a variety of quantum-enhanced communication, sensing, and computing capabilities. Entanglement-assisted communication (EACOMM) leverages entanglement pre-shared by communication parties to boost the rate of classical information transmission. Pioneering theory works showed that EACOMM can enable a communication rate well beyond the ultimate classical capacity of optical communications, but an experimental demonstration of any EACOMM advantage remains elusive. Here, we report the implementation of EACOMM surpassing the classical capacity over lossy and noisy bosonic channels. We construct a high-efficiency entanglement source and a phase-conjugate quantum receiver to reap the benefit of pre-shared entanglement, despite entanglement being broken by channel loss and noise. We show that EACOMM beats the Holevo-Schumacher-Westmoreland capacity of classical communication by up to 14.6\%, when both protocols are subject to the same power constraint at the transmitter. As a practical performance benchmark, a classical communication protocol without entanglement assistance is implemented, showing that EACOMM can reduce the bit-error rate by up to 69\% over the same bosonic channel. Our work \ZZ{opens a route to provable quantum advantages in a wide range of quantum information processing tasks.}

\end{abstract}

\maketitle
{\em Introduction.---}Entanglement as a nonclassical resource is the cornerstone for a wide range of quantum information processing (QIP) applications including quantum-secured communication~\cite{ekert1991}, quantum-enhanced sensing~\cite{Giovannetti_2001}, and quantum computing~\cite{shor1999polynomial}. In addition, entanglement pre-shared by communication parities can increase the rate of transmitting classical information, a paradigm known as entanglement-assisted (EA) communication (EACOMM)~\cite{bennett1999entanglement,bennett2002entanglement,holevo02,shor2004classical,hsieh2008entanglement,zhuang2017additive,wilde2012quantum,wilde2012information}. \ZZ{The pioneering work by Bennett, Shor, Smolin, and Thapliyal~\cite{bennett2002entanglement} showed that the channel capacity with EA surpasses the ultimate classical capacity without EA, i.e., the Holevo-Schumacher-Westmoreland (HSW) capacity~\cite{hausladen1996classical,schumacher1997sending,holevo1998capacity}. Surprisingly, for lossy and noisy bosonic channels, which are ubiquitous in optical and microwave communications, photonic sensing, and one-way quantum computing~\cite{menicucci2006universal}, the ratio between the EA capacity and the HSW capacity can diverge. Notably, the EA-capacity advantage sustains even if a lossy and noisy channel breaks the initial pre-shared entanglement.}

This seminal EA-capacity result, albeit encouraging, does not elucidate an EACOMM protocol to reap the promised advantage. \ZZ{In this regard, superdense coding is a well-studied EACOMM scenario that leverages stronger-than-classical correlations between entangled photons to encode more than one classical bit of information on each transmitted photon~\cite{bennett1992,guo2019advances}. However, EACOMM experiments~\cite{mattle1996dense,barreiro2008beating,schaetz2004quantum,prevedel_2011,chiuri_2013,williams2017superdense,liu2016efficient,hu2018beating} based on the polarization, temporal, and path degrees of freedom have dismissed the phase correlations embedded in entangled bosonic modes, thereby unable to beat the HSW capacity.} Studies on EACOMM protocols over bosonic channels encompassed continuous-variable superdense coding~\cite{ban1999quantum,braunstein2000,ban2000quantum} and mode permutations or selections encoding~\cite{wilde2012information,anshu2019building,qi2018applications,khabbazi2019union}. Unfortunately, the former failed to surpass the HSW capacity due to poor performance in the presence of channel loss and noise~\cite{sohma2003,mizuno2005experimental,barzanjeh2013,li2002quantum}, whereas the latter requires large-volume quantum memories that are not yet available. Recently, Ref.~\cite{shi2020practical} formulated a theoretical framework to devise the optimal entangled state and encoding format suitable for loss and noise resilient EACOMM. The theory work also proposed practical quantum receiver structures to enable an EACOMM rate superior to the HSW capacity. 

Here, we report an EACOMM experiment over lossy and noisy bosonic channels at communication rates up to 14.6\%$\pm$6.5\% above the HSW capacity. \ZZ{In contrast to many superdense coding protocols that resort to the probabilistic arrival of single photons at the receiver due to channel loss, our EACOMM protocol harnesses the phase correlations between entangled bosonic modes so that all transmitted bits are decoded. A low-gain quantum phase-conjugate receiver (PCR) is constructed to effectively reduce the noise power while preserving the phase correlations.} Apart from benchmarking against the ultimate HSW capacity, we show that EACOMM achieves error probabilities up to 69\% lower than what a practical CCOMM system can afford. \ZZ{Our work achieves a provable quantum advantage and would create new opportunities for entanglement-enhanced QIP.}

\begin{figure}
    \centering
    \includegraphics[width=.48\textwidth]{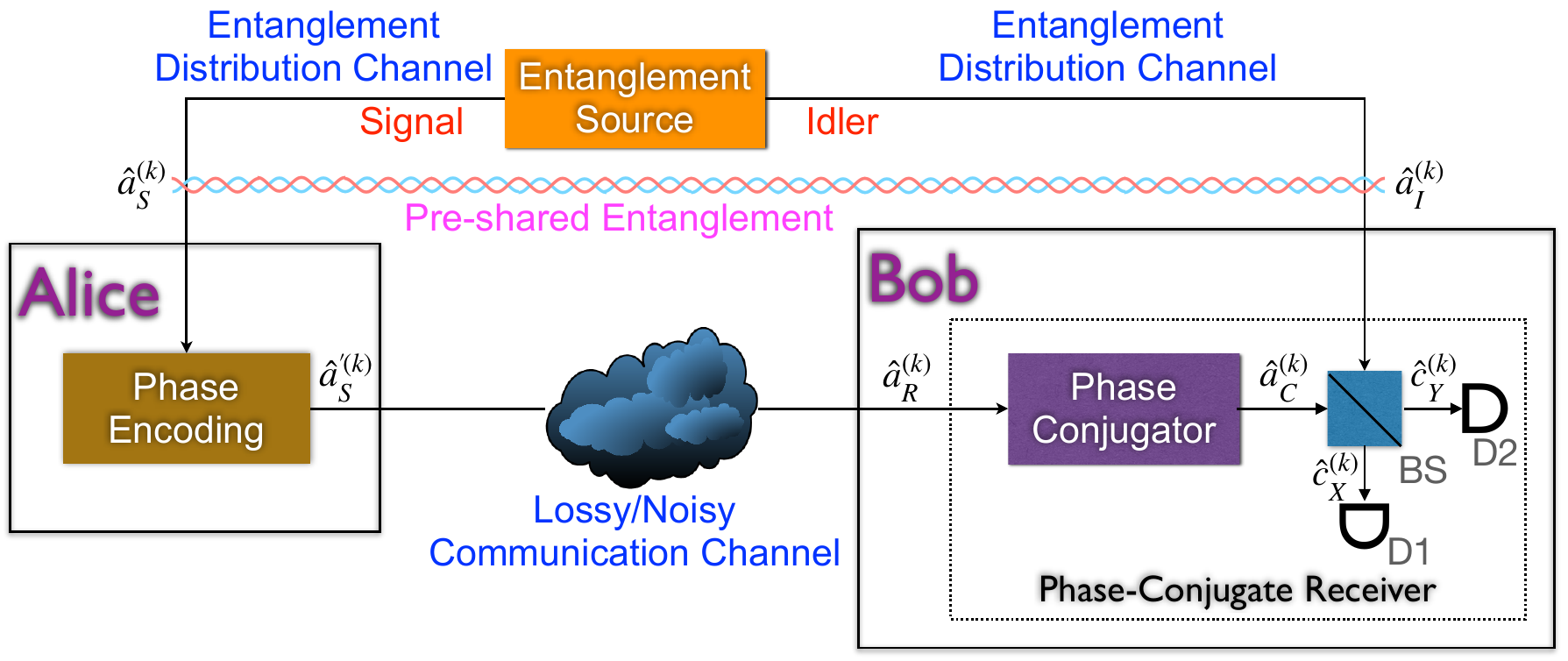}
    \caption{Schematic of the EACOMM protocol. An entanglement source distributes two-mode squeezed vacuum signal-idler pairs $\{\hat{a}_S^{(k)},\hat{a}_I^{(k)}\}$ to Alice and Bob. Alice phase encodes on $\hat{a}_S^{(k)}$ and transmits the \ZZ{encoded} modes $\hat{a}_S^{\prime(k)}$ to Bob through a lossy and noisy channel. Bob produces $\hat{a}_C^{(k)}$ by phase conjugating the received modes $\hat{a}_R^{(k)}$. $\hat{a}_C^{(k)}$ and $\hat{a}_I^{(k)}$ interfere on a balanced beam splitter (BS), whose two output arms are counted by two photodetectors D1 and D2 to derive the difference photon number, from which Alice's encoded classical bit is inferred. See text for details of the evolution of mode operators.}
    \label{fig:EACOMM_protocol}
\end{figure}

{\em Protocol.---}The schematic for the implemented EACOMM protocol is sketched in Fig.~\ref{fig:EACOMM_protocol}.
Key components include 1) an entanglement source; 2) two low-loss entanglement distribution channels connecting the source to Alice and to Bob; 3) phase encoding on Alice's share of the entanglement, i.e., the signal; and 4) a PCR that Bob operates to perform joint measurements on the received signal from a lossy and noisy channel and Bob's share of the entanglement, i.e., the idler.

Prior to transmitting one bit of classical information, the entanglement source emits $M$ i.i.d. two-mode squeezed vacuum (TMSV) signal-idler mode pairs, described by annihilation operators $\left\{\hat{a}_S^{(k)}, \hat{a}_I^{(k)}\right\}_{k = 1}^M$, and sends the signal modes to Alice and the idler modes to Bob through two low-loss, noiseless entanglement-distribution channels. The mean photon number of a signal or an idler mode is defined as $N_S \equiv \left\langle \hat{a}_S^{\dag (k)}\hat{a}_S^{(k)}\right\rangle = \left\langle \hat{a}_I^{\dag (k)}\hat{a}_I^{(k)}\right\rangle$. To encode a classical bit $b \in \{0,1\}$, Alice applies binary phase-shift keying on {\em all} $M$ signal modes, yielding encoded signal modes $\left\{\hat{a}_S^{\prime(k)} = (-1)^b \hat{a}_S^{(k)}\right\}_{k = 1}^M$ that are subsequently transmitted to Bob through a bosonic thermal-loss channel~\cite{Weedbrook_2012}, $\mathcal{L}^{\kappa, N_B}$, characterized by the transmissivity $\kappa$ and a per-mode noise photon number $N_B$. The noise photons are effectively introduced by thermal background modes $\left\{\hat{a}_B^{(k)}\right\}_{k = 1}^M$, each with a mean photon number of $\left\langle\hat{a}_B^{\dag(k)} \hat{a}_B^{(k)} \right\rangle = N_B/(1-\kappa)$. \ZZ{The mode evolution relation in the Heisenberg picture gives Bob's received signal modes $\left\{\hat{a}_R^{(k)} = \sqrt{\kappa}\hat{a}_S^{\prime(k)} + \sqrt{1-\kappa}\hat{a}_B^{(k)}\right\}_{k = 1}^M$ that contain $N_B$ thermal noise photons per mode.} \ZZ{Both the employed TMSV state and phase encoding have been proven optimum for EACOMM~\cite{shi2020practical}.}

To decode the classical bit, Bob uses a PCR to perform a joint measurement on the received signal modes $\left\{\hat{a}_R^{(k)}\right\}_{k = 1}^M$ and idler modes $\left\{\hat{a}_I^{(k)}\right\}_{k = 1}^M$ from the entanglement source~\cite{Guha2009}. In the PCR, phase-conjugate modes, $\hat{a}_C^{(k)}$, of the received signal are obtained in a parametric process with gain $G$, viz. $\left\{\hat{a}_C^{(k)} = \sqrt{G}\hat{a}_v^{(k)} + \sqrt{G-1} \hat{a}_R^{\dag (k)}\right\}_{k = 1}^M$, where $\left\{\hat{a}_v^{(k)}\right\}_{k = 1}^M$ are vacuum modes. The phase-conjugate modes then interfere with the idler modes on a balanced beam splitter, leading to the modes $\left\{\hat{c}_X^{(k)} = \left(\hat{a}_C^{(k)} + \hat{a}_I^{(k)}\right)/\sqrt{2}\right\}_{k = 1}^M$ and $\left\{\hat{c}_Y^{(k)} = \left(\hat{a}_I^{(k)}-\hat{a}_C^{(k)}\right)/\sqrt{2}\right\}_{k = 1}^M$ at the two output ports. Photon counting at each output port \ZZ{measures $M$ modes, so the two detectors generate two jointly Gaussian variables $N_X,N_Y$ in the asymptotic limit of $M \gg 1$}. \ZZ{The difference photon number, defined as $N \equiv N_X - N_Y$, is dependent on the phase-insensitive cross correlations $\left\{\left\langle \hat{a}^{\dag(k)}_C\hat{a}^{(k)}_I\right\rangle\right\}_{k = 1}^M$, which stem from the phase-sensitive cross correlations $\left\{\left\langle\hat{a}^{(k)}_S\hat{a}^{(k)}_I \right\rangle\right\}_{k = 1}^M$ of the TMSV states.} The decoded classical bit $\tilde{b}$ is set to 0 (1) when $N \geq 0$ ($N < 0$). The bit-error rate (BER) of EACOMM using TMSV states and the PCR can be analytically derived as~\cite{shi2020practical}
\begin{align}
P_e= \frac{1}{2}{\rm erfc}\left(\sqrt{ \frac{2M \eta_D\kappa_I\kappa N_S(N_S+1)}{N_B(1+2\delta\eta+2\eta_D\kappa_IN_S)}}\right), 
\label{PE_PCR}
\end{align}
in the $N_B\gg1, M\gg1$ limit (see~\cite{supp} for the full formula), where $\eta_D$ is the effective detection efficiency, $\kappa_I$ is idler's overall efficiency including the source and entanglement-distribution efficiencies, and $\delta\eta$ models deviation of the BS transmissivity from 50\%.

With equal probability of Alice sending zeros and ones, \ZZ{the BER then determines the mutual information between Alice and Bob, obtained by transmitting $M$ modes},  as 
\be
\label{eq:mutual_info}
I(A;B) = 1+P_e\log_2(P_e)+(1-P_e)\log_2(1-P_e).
\ee 
Without EA, the HSW capacity per mode, subject to the same mean photon-number constraint $N_S$, has been derived as~\cite{GiovannettiV2014}
\begin{equation}
    C(\mathcal{L}^{\kappa,N_B}) = g(\kappa N_S + N_B) - g(N_B),
\end{equation}
where $g(N) = (N+1)\log_2(N+1) - N\log_2(N)$ is the entropy of a thermal state with mean photon number $N$. Demonstrating $I(A;B) > M C(\mathcal{L}^{\kappa,N_B})$ will prove that EACOMM surpasses the ultimate classical capacity. 

\begin{figure*}[hbt]
    \centering
    \includegraphics[width=1\textwidth]{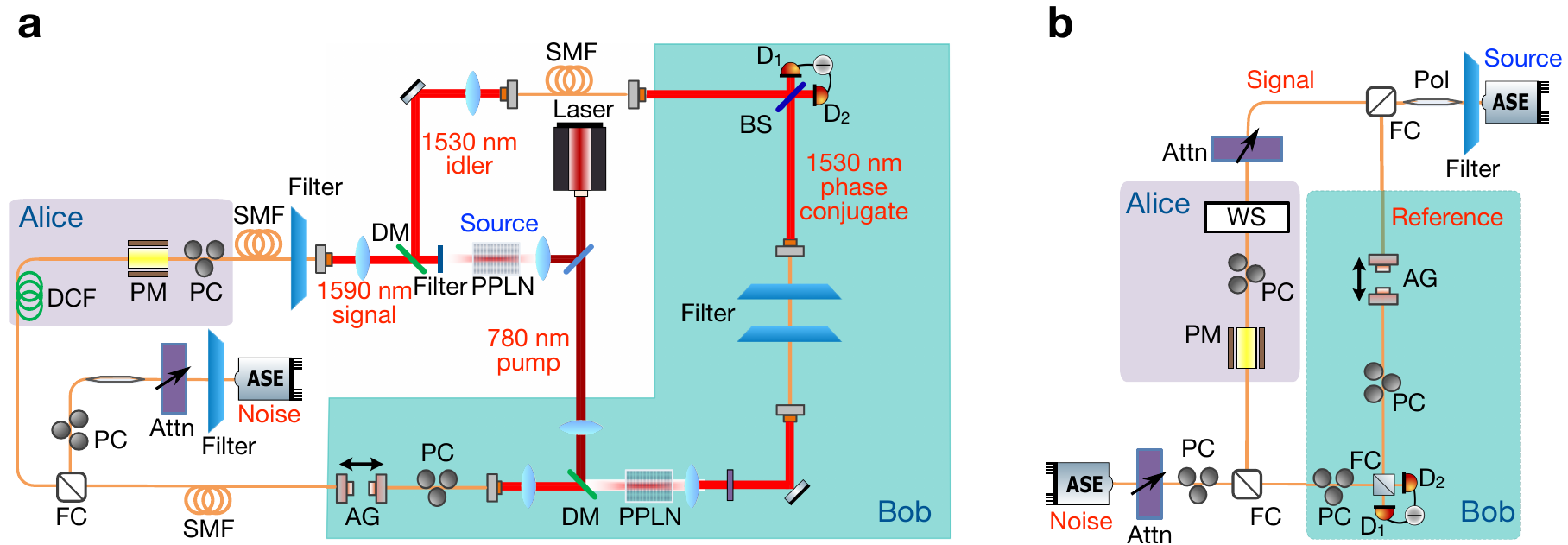}
    \caption{(a) Experimental diagrams for (a) EACOMM and (b) CCOMM. AG: air gap; ASE: amplified spontaneous emission; Attn: attenuator; BS: beam splitter; D: detector; DCF: dispersion compensating fiber; DM: dichroic mirror; FC: fiber coupler; PC: poloarization controller; PM: phase modulator; Pol: polarizer; PPLN: periodically-poled lithium niobate; SMF: single-mode fiber; WS: waveshaper.
    }
    \label{fig:exp_diagram}
\end{figure*}

{\em Experimental setup.---}The experimental diagram for EACOMM is depicted in Fig.~\ref{fig:exp_diagram}(a). The entanglement source comprises a periodically-poled lithium niobate (PPLN) crystal pumped by a 780-nm laser to produce broadband spontaneous parametric down conversion (SPDC) photons centered at 1560 nm. In the experiment, we pick the signal and idler modes to situate, respectively, around 1590 nm and 1530 nm. Due to energy conservation in SPDC, the signal and idler form entangled mode pairs each described by a TMSV state. A dichroic mirror separates the signal and idler modes. The signal and idler are subsequently coupled into single-mode fibers through two collimators. A flat-top optical filter is then applied on the signal to carve out a 16-nm band centered around 1590 nm, corresponding to an optical bandwidth of $W \sim 2$ THz. The signal photons are distributed to Alice while the idler photons are sent to Bob through two entanglement-distribution channels constituted of low-loss single-mode fibers. The overall efficiency $\kappa_I$ for the idler distribution and storage is measured to be 96\%.

To encode a classical bit $b$ at Alice's transmitter, an electro-optic modulator (EOM) driven by a BER tester imparts a $T$-second-long phase shift of $b\pi$ on $M = WT$ signal modes. The phase modulated signal modes are sent to Bob through an optical fiber link. \ZZ{An L-band amplified spontaneous emission (ASE) source, filtered to a 16-nm band centered around 1590 nm, serves as the thermal light source due to its second-order coherence property~\cite{doronin2019second} and multimode photon statistics~\cite{wong1998photon}.} The ASE light is combined with the encoded signal on a fiber coupler. We construct a free-space air gap to fine tune the relative delay between the signal and idler photons so that they simultaneously arrive at the PCR.

At Bob's terminal, we couple the signal photons from fiber to free space via a collimator. The signal is then sent to a second PPLN crystal pumped by a 780-nm laser to generate the phase-conjugate modes at the idler wavelength of 1530 nm via a \ZZ{difference-frequency generation} process with gain $G= 1+ 0.257\times 10^{-3}$. The output of the PPLN crystal is coupled back to optical fibers via a collimator. Two cascaded bandpass filters then reject the signal photons at 1590 nm, and the remaining phase-conjugate photons are coupled back to free space. The phase-conjugate photons interfere with the idler photons on a 50:50 beam splitter whose $\delta \eta \sim 10^{-3}$. The photons at the two output ports of the beam splitter are diverted to a balanced detector with an effective detection efficiency of $\eta_D=95\%$, which includes the 99\% quantum efficiency of the photodiodes and the interference visibility of 98\%. \ZZ{Note that the measurement is not based on either coincidence counting or Hong-Ou-Mandel interference because at the receiver the noise photons are more than 8 orders of magnitude brighter than the photons originating from the source.} The output electrical signal from the detector is directed to the BER tester.

\begin{figure}
    \centering
    \includegraphics[width=.45\textwidth]{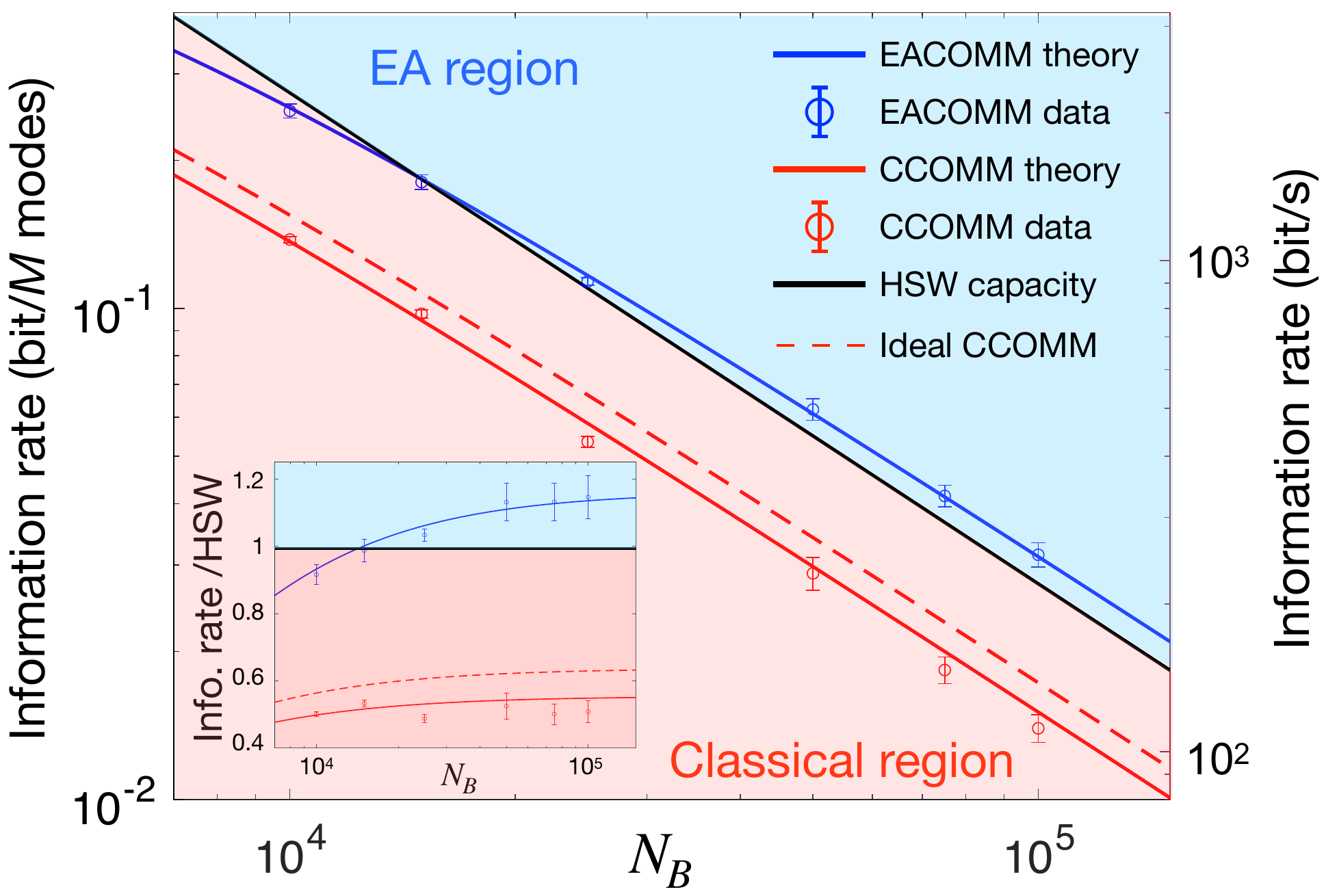}
    \caption{The information rate for EACOMM (blue), practical CCOMM (red), and the classical capacity (black) at different thermal background noise levels. Curves: theory; dots: experimental data. (Inset) EACOMM rate over the HSW capacity (blue) and the practical CCOMM rate over the HSW capacity (red). Dashed curves: theory for ideal CCOMM with $\kappa_F = 1$. Blue shaded area: EA region. Red shaded area: classical region. Error bars derived from 15 consecutive BER measurements each using $10^{4}$ bits. In the experiment, $\kappa=0.014$, $M= 2.5\times 10^{8}$, $\ZZ{\kappa_I = 0.96, \eta_D = 0.95, \delta\eta = 10^{-3}}$, and $N_S = 7.8\times 10^{-4}$.
    }
    \label{fig:rate_NB}
\end{figure}

{\em Demonstrating quantum advantages.---}We first demonstrate that EACOMM over lossy and noisy channels can achieve a rate higher than any CCOMM protocol without EA can afford, thereby proving EACOMM's quantum advantage. In the experiment, the power of the transmitted signal is fixed at $P_S = 195$ pW so that $N_S = P_S/\hbar\omega_0 W =7.8\times 10^{-4}$, where $\hbar$ is the reduced Planck constant, and $\omega_0$ is the frequency of the signal photons. In measuring the BERs, $N_B$ is tuned from $10^4$ to $10^5$ by cranking up the output power of the ASE source. The corresponding mutual information given by Eq.~\eqref{eq:mutual_info} is plotted alongside the HSW capacity in Fig.~\ref{fig:rate_NB}, showing experimental EACOMM's advantages at $N_B > 5\times 10^4$. As we see, the theory (blue curve) agrees well with the experiment results (blue dots); the disadvantage at low $N_B$ is due to effects from additional loss in the receiver~\cite{supp}. This result indicates that the EACOMM's advantage becomes even more pronounced over a more noisy channel. 

Given the optical bandwidth $W$ and the source brightness $N_S$, the HSW capacity sets an ultimate limit for the communication rate without EA. In practice, however, approaching the classical capacity would require the optimal encoding and quantum measurements on each signal mode, which would be beyond the reach of current technology. To experimentally assess how practical CCOMM without EA performs, we implement a protocol based on broadband light and multimode encoding and measurements, as illustrated in Fig.~\ref{fig:exp_diagram}(b). Broadband light was previously utilized by floodlight quantum key distribution to boost the secret-key rates~\cite{zhuang2016floodlight,zhang2017floodlight,zhang2018experimental}. In the CCOMM experiment, ASE light is filtered to 16-nm bandwidth and then split into two arms that differ substantially in the optical power levels. The weak output arm with a per-mode mean photon number $N_S \ll 1$ serves as the signal and is distributed to Alice, whereas the strong output arm with a per-mode mean photon number $N_R \gg 1$ becomes a broadband reference and is sent to Bob. From Alice's perspective, her received quantum states are identical to the marginal entangled state in EACOMM after tracing out the idler modes. As such, we make use of the same phase-modulation scheme to encode classical bits, as what the EACOMM protocol adopts. At Bob's terminal, the received signal and the reference interfere on a 50:50 fiber coupler, whose two output arms are measured by a balanced detector that produces a difference photocurrent. Like the EACOMM experiment, a phase-locking servo loop is implemented to ensure stable BER measurements. 
Given $N_R\gg 1$ and $N_B\gg N_S$, the error rate of the broadband light homodyne detection approaches the homodyne detection on coherent states~\cite{supp}
\be 
\label{eq:PE_CCOMM}
P_e=\frac{1}{2}\text{erfc} \left(\sqrt{\frac{ M \kappa\kappa_F N_S}{N_B+1/2}}\right),
\ee
where $\kappa_F=0.87$ is a fitting parameter accounting for experimental nonidealities including imperfect dispersion matching between the signal and the reference and detector balancing.
\begin{figure}[hbt]
    \centering
    \includegraphics[width=.45\textwidth]{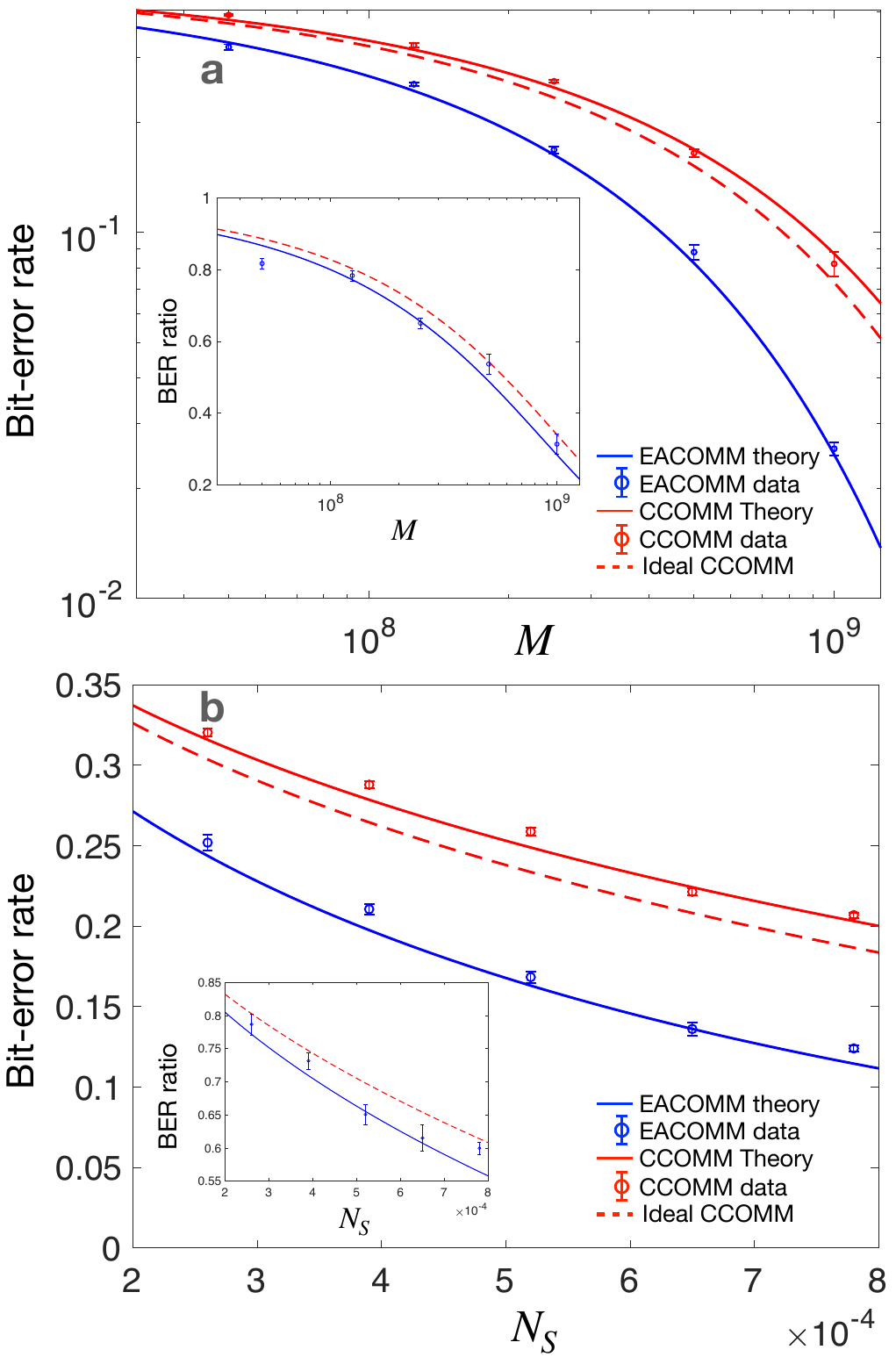}
    \caption{(a) Bit-error rate vs number of modes per encoding ($M$). In the measurements, $\kappa=0.043, N_B= 2.1\times 10^{4}$, and $N_S= 5.2\times 10^{-4}$. (b) Bit-error rate vs the source brightness ($N_S$). In the measurements, $\kappa=0.043, M= 2.5\times 10^{8}, \ZZ{N_R = 1.2\times 10^{3},}$ and $N_B= 2.1\times 10^{4}$. (Insets) BER ratio between EACOMM and CCOMM. Dots: experimental data; curves: theory; dashed curves: ideal CCOMM with $\kappa_F = 1$.Error bars derived from 15 consecutive BER measurements each based on $10^{4}$ bits.}
    \label{fig:BER}
\end{figure}

The performance of EACOMM is compared with that of CCOMM under three parameter settings. First, the BERs of the CCOMM protocol under different amount of channel background thermal noise are measured. The corresponding information rates are plotted in Fig.~\ref{fig:rate_NB}, showing good agreement with the theoretical model. EACOMM enjoys an up to 126\% information-rate advantage over the CCOMM protocol. We then measure the BERs of the EACOMM and CCOMM protocols at different number of modes per encoding, i.e., the encoding rate, and plot the experimental data in Fig.~\ref{fig:BER}(a), showing a substantial BER advantage for EACOMM over CCOMM. Fig.~\ref{fig:BER}(b) plots the BER data taken at different source brightness $N_S$. EACOMM demonstrates a reduced BER at all $N_S$ levels, with the largest BER reduction of 40\% measured at $N_S = 7.8\times 10^{-4}$. 

{\em Discussions.---} \ZZ{EACOMM uses pre-shared entanglement to improve the rates of transferring {\em classical information}, instead of quantum bits~\cite{cuevas2017experimental}. The pre-shared entanglement would be more efficiently distributed at the optical wavelengths~\cite{yin2017satellite}. The signal photons can then be frequency converted~\cite{lambert2020coherent} to support microwave EACOMM.}

Quantum illumination (QI)~\cite{shapiro2020quantum} also uses entanglement over lossy and noisy bosonic channels to detect the presence of a target~\cite{Tan2008,barzanjeh2015microwave,lopaeva2013experimental,zhang2015,barzanjeh2020microwave} or to defeat a passive eavesdropper~\cite{Shapiro2009,zhang2013,shapiro2014secure}. None of the previous QI experiments~\cite{zhang2013,zhang2015} is capable of beating the HSW capacity. Specifically, the use of an optical amplifier in QI secure communication breaks the pre-shared entanglement even before sending the encoded signal and thus forfeits the benefit of EACOMM. Also, the optical parametric amplifier receiver (OPAR) used in the previous QI experiments has a simple configuration due to the simultaneously interacting signal, idler, and pump on a nonlinear crystal, but this setup introduces additional loss on the idler beam such that EACOMM's stringent requirements on the efficiency of the quantum receiver cannot be satisfied. The PCR, in contrast, first generates a {\em bright} phase-conjugate beam of the signal so that any additional loss nearly has no affect on the receiver performance. As a consequence, the PCR is able to attain a large EACOMM advantage sufficient to outperform the classical capacity and is also envisaged to substantially improve the QI target detection and secure communication based on OPAR. \ZZ{Moreover, the EACOMM advantage can be extended and generalized to a scenario without a phase reference~\cite{zhuang2021} and a quantum network with multi-access channels.}

Although we have demonstrated EACOMM surpassing the HSW capacity, the current implementation based on the PCR does not saturate the EA capacity. A recent study proposed a quantum-receiver structure based on single-photon-level sum-frequency generation~\cite{zhuang2017} and multimode interference to achieve the $\log_2(N_S)$ scaling of EACOMM~\cite{guha2020infinite}, thereby pointing to a promising route towards realizing a larger EACOMM advantage over CCOMM.

{\em Conclusions.---}We have developed an efficient entanglement source and quantum receiver to demonstrate EACOMM beyond the classical capacity. Our work demonstrates the power of pre-shared entanglement in enhancing the rate of transmitting classical information over lossy and noisy bosonic channels. This result would pave a new avenue toward utilizing entanglement to achieve a provable quantum advantage in applications involving substantial loss and noise, such as low probability of intercept~\cite{bash2015quantum,shapiro2019quantum}, covert sensing~\cite{gagatsos2019covert}, and non-invasive imaging~\cite{taylor2013biological}.

\section*{Acknowledgments}
We gratefully acknowledge funding support by the National Science Foundation Grant No.~CCF-1907918, ECCS-1828132, EEC-1941583, and General Dynamics Mission Systems. QZ also acknowledges support from Defense Advanced Research Projects Agency (DARPA) under Young Faculty Award (YFA) Grant No.~N660012014029. The authors thank HC Photonics for providing the nonlinear crystals, Jeffrey Shapiro for valuable comments on the manuscript, and William Clark and Saikat Guha for helpful discussions.

\bibliography{myref}



\section*{Supplementary material}

# Supplemental Material
# Entanglement-Assisted Communication Surpassing the Ultimate Classical Capacity

Shuhong Hao,[1] Haowei Shi,[2] Wei Li,[1] Quntao Zhuang,[3, 2] and Zheshen Zhang[1, 2, *]

[1]*Department of Materials Science and Engineering,
University of Arizona, Tucson, Arizona 85721, USA*
[2]*James C. Wyant College of Optical Sciences, University of Arizona, Tucson, Arizona 85721, USA*
[3]*Department of Electrical and Computer Engineering,
University of Arizona, Tucson, Arizona 85721, USA*

## I. EXPERIMENTAL DETAILS

In the entanglement-assisted communication (EACOMM) experiment, the entanglement source comprises a 25-mm-long type-0 periodically-poled lithium niobate (PPLN) crystal (HC Photonics) embedded in an oven temperature stabilized at 115 Celsius degree. To reduce the loss penalty on the generated TMSV states due to mismatched signal and idler collecting spatial modes, the pump laser is loosely focused onto the crystal to suppress the SPDC photons emitted into higher-order spatial modes. In addition, the focal length of the lens for the signal (idler) after the crystal is optimized to be 200 mm (300 mm), leading to a collecting diameter of 0.316 mm (0.206 mm) in the crystal. In doing so, additional loss on the idler modes is minimized [1–3], a key to achieve an EACOMM advantage over the classical capacity. The heralding efficiency of the idler photon conditioned on detecting a signal photon in a low pump power situation is estimated to be ∼ 99%. The signal and idler beams are separated by a long-pass dichroic mirror with a cutoff wavelength at 1550 nm and then coupled into single-mode fibers through two collimators (Thorlabs F240FC/F260FC). The input face of the idler fiber patch cable is anti-reflection (AR) coated to minimize extra loss. The signal is first filtered by a 16-nm flat-top optical filter centered at 1590 nm and then sent to an electro-optic phase modulator (PM) with a built-in polarizer (Thorlabs LN65S). The input polarization to the PM is controlled by a paddle to maximize the transmission. Due to the broadband signal and idler, group-velocity dispersion (GVD) induced by the optical fiber would reduce the performance of the quantum phase-conjugate receiver (PCR). A regular approach to overcome GVD is to place dispersion compensating fibers (DCFs) on both the signal and the idler. However, the additional loss that DCF introduces on the idler would weaken the quantum advantage of EACOMM. To mitigate this challenge, we *overcompensate* the GVD on the signal by adding 8.3-meter-long DCFs with a dispersion parameter of -90.4 ps/nm/km while leaving the idler in single-mode fibers. In doing so, we effectively leverage the phenomenon known as nonlocal dispersion cancellation [4, 5] to reinstate a near-optimum performance of the PCR. An amplified spontaneous emission (ASE) source followed by a 16-nm flat-top optical filter centered at 1590 nm produces broadband light to emulate the channel thermal noise. The ASE light was shown to have second-order coherence property [6] and photon statistics close to those of the multimode thermal state [7]. An ASE source is chosen over a light lamp to produce sufficient power at the telecommunication wavelength for our experiment. A polarizer rejects one polarization, and the rest of the broadband light is polarization controlled and mixed with the encoded signal on a fiber coupler. The splitting ratio of fiber coupler is chosen as 90:10 under $M$ or $N_s$ measurements and as 30:70 under $N_B$ measurements. An air gap (AG) subsequently fine tunes the relative delay between the signal and the idler. At the PCR, the heralding efficiency conditioned on detecting a signal photon is estimated to be ∼ 95% when the thermal noise injection is switched off. After polarization controlled by a paddle, the signal is coupled back to free space via a collimator (Thorlabs F240FC). In free space, a DM combines the signal with the pump. The combined beam is injected into a second PPLN crystal temperature stabilized at 116 °C to generate the phase-conjugate beam at 1530 nm. The phase-conjugate beam is coupled into single-mode fiber via a collimator (Thorlabs F240FC) and then filtered by two 16-nm flat-top optical filters centered at 1530 nm. The filtered phase-conjugate beam and the idler are both coupled to free space through AR-coated patch cables and collimators (Thorlabs F240FC) and interfere with each other on a 50:50 beam splitter (BS) cube. The visibility of the interference is optimized to >98% using a lens on the

---

[*] zsz@arizona.edu

phase-conjugate beam. The beams of the two output ports of the BS are focused onto a homemade balanced detector comprising two photodiodes both with a 99% quantum efficiency (Laser Components, InGaAs 1550). The difference photocurrent is amplified by a transimpedance amplifier (TIA) with a gain of $5 \times 10^7$ V/A (Femto LCA-100k-50M). The voltage signal is filtered by an electrical low-pass filter to reject out-of-band noise and is then split into two arms, one going to a lock-in amplifier while the other going to the bit-error rate (BER) tester. The output of the lock-in amplifier is further processed by a proportional–integral–derivative (PID) controller to generate an error signal that is combined with the dither signal from the same lock-in amplifier to feed to the PM for the implementation of a servo loop that locks the relative phase between the phase-conjugate and idler beams. The readings of the BER tester is recorded under different experimental settings.

In the classical-communication experiment, the output from an ASE source is filtered by a flat-top 16-nm optical filter centered at 1550 nm. A polarizer then rejects one polarization while passing the other. Then, a 99:1 unbalanced fiber coupler (FC) produces a weak broadband signal with a per-mode mean photon number $N_S \ll 1$ and a bright broadband reference with a per-mode mean photon number $N_R \gg 1$. A tunable fiber attenuator further reduces the signal power to a level set by the requirements in different experimental runs. A waveshaper (Finisar 1000A) is applied on the signal to compensate for its dispersion disparity with the reference. A PM driven by the BER tester subsequently encodes on the signal. The noisy channel is emulated by injecting ASE light with the same optical bandwidth as the signal. An attenuator is applied on the ASE noise to control the noise power of the channel. A polarization controller (PC) ensures that the ASE noise and the signal share the same polarization. At Bob's receiver, the propagation time of the reference is fine tuned by an AG so that it can efficiently interfere with the signal on a 50:50 FC. Prior to the interference, the polarizations of the signal and the reference are controlled by two PCs on both arms. The two output ports of the FC are directed and measured by a balanced detector (Thorlabs PDB450C) with a 80% quantum efficiency. The difference photocurrent is amplified by a TIA with gain $10^6$ V/A. The voltage signal is filtered by a low-pass filter. A portion of the filtered voltage signal is diverted to a lock-in amplifier followed by a PID controller to implement a servo loop that locks the relative phase between the signal and the reference, akin to the EACOMM experiment. The rest of the voltage signal is measured by the BER tester.

## II. THEORETICAL MODEL FOR PHASE-CONJUGATE RECEIVER

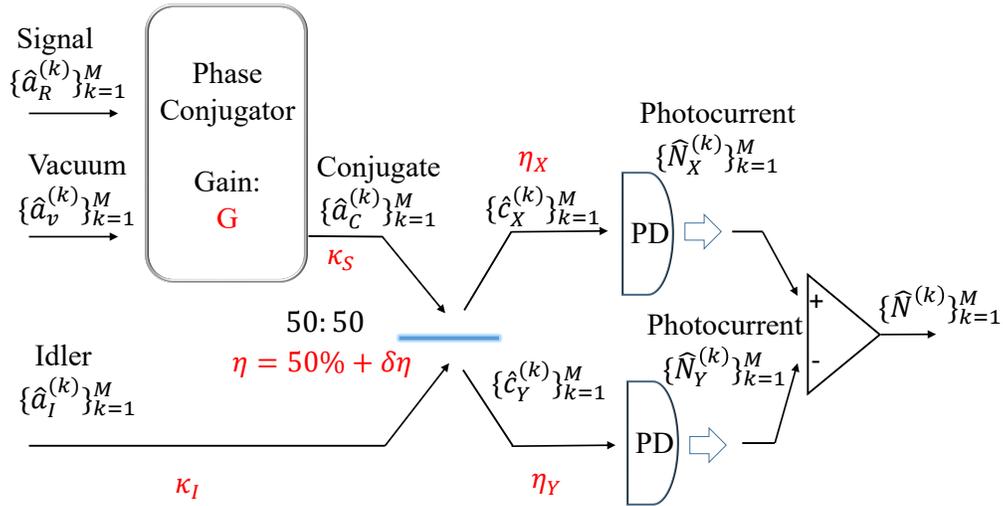

Figure 1. The schematic of PCR, with experimental imperfections highlighted in red.

We formulate a complete theoretical model, based on Ref. [8], to analyze the BER in the intermediate region where $G - 1$ is not sufficiently large. In this region, higher-order terms of $N_S$ cannot be ignored. The model also accounts for experimental imperfections including the transmissivity after the phase conjugator ($\kappa_S$), the transmissivity of the

entanglement distribution channel for the idler ($\kappa_I$), the deviation $\delta\eta$ of the transmissivity $\eta$ from 50% for the BS that interferes the phase-conjugate and idler beams, and the detector quantum efficiency $\eta_X = \eta_Y = \eta_D$, as shown in Fig. 1.

Considering all nonidealities, the mean $\mu_\pm$ of measurement result $\hat{N}$ conditioned on the phase modulation $\pm 1$ can be calculated and the difference

$$\mu_+ - \mu_- = 4C_{CI}\eta_D \sqrt{1 - 4(\delta\eta)^2} . \tag{1}$$

When the beamsplitter is balanced, the variances are given by

$$\sigma_\pm^2|_{\eta=1/2} = \eta_D N_I + 2\eta_D^2 N_C N_I + \eta_D N_C + 2\eta_D^2 C_{CI}^2 \sim \eta_D N_C, \tag{2}$$

while the deviation of the variances for non-zero $\delta\eta$ is given as

$$\begin{aligned}\delta\sigma_\pm^2 &\equiv \sigma_\pm^2 - (\sigma_\pm^2|_{\eta=1/2}) = \eta_D^2[-16C_{CI}^2(\delta\eta)^2 + 4(\delta\eta)^2(N_C - N_I)^2 \pm 8C_{CI}\delta\eta \sqrt{1 - 4(\delta\eta)^2}(N_C - N_I)] \\ &\sim \eta_D^2 N_C[\pm 8C_{CI}\delta\eta + 4(\delta\eta)^2 N_C],\end{aligned} \tag{3}$$

where $N_C = (G - 1)\kappa_S(\kappa N_S + N_B + 1), N_I = \kappa_I N_S, C_{CI} = C_p \sqrt{(G - 1)\kappa_I \kappa_S}$, with $C_p = \sqrt{\kappa N_S(1 + N_S)}$. We find that the effect of an unbalanced $\eta$, i.e., $\delta\eta > 0$, is negligible at $(\delta\eta)^2 \ll 1$ (so that the mean deviation is small) and $\eta_D[8C_{CI}\delta\eta + 4(\delta\eta)^2(G-1)N_B] \ll 1$ (so that $\delta\sigma_\pm^2 \ll \sigma_\pm^2|_{\eta=1/2}$), which is in accordance with the operational regime of the experiment. Now let $\eta = 1/2$, define the exponential decay rate $R$ of the error rate $P_e \sim \exp(-MR)/2$ as the error exponent. The exact error exponent is given by

$$R = \frac{2\eta_D^2(G - 1)\kappa\kappa_I\kappa_S N_S(N_S + 1)}{\eta_D(N_I + N_C) + \eta_D^2\left(2N_C N_I + 2C_{CI}^2\right)}, \tag{4}$$

For $N_B \gg 1$ and $N_S \ll N_B$, we take the approximation of $N_C \simeq (G - 1)\kappa_S N_B$, and then

$$\begin{aligned}R &\simeq \frac{2\eta_D^2(G - 1)\kappa\kappa_I\kappa_S N_S(N_S + 1)}{\eta_D[(G - 1)\kappa_S N_B + \kappa_I N_S] + 2\eta_D^2(G - 1)\kappa_I N_S \kappa_S N_B} \\ &= \frac{2\eta_D \kappa_I \kappa N_S(N_S + 1)}{N_B(1 + 2\eta_D \kappa_I N_S) + \kappa_I N_S/[(G - 1)\kappa_S]}.\end{aligned} \tag{5}$$

One sees that the first-order term $\kappa_I N_S/[(G - 1)\kappa_S]$ in the denominator will no longer be negligible if the gain $G$ of the phase conjugator falls near 1, which will significantly undermine the quantum advantage. To suppress this term, we require

$$(G - 1)\kappa_S N_B \gg \kappa_I N_S . \tag{6}$$

In conclusion, we expect a sufficiently large gain $G$, according to Eq. (6), to guarantee a quantum advantage for EACOMM. In this case, $\kappa_S$ does not influence the leading term in the error exponent, so it becomes

$$R \simeq \frac{2\eta_D \kappa_I \kappa N_S(N_S + 1)}{N_B(1 + 2\eta_D \kappa_I N_S)}. \tag{7}$$

The BER formula in Eq. (??) of the main paper is derived from the asymptotic result in Eq. (7), but the exact formula in Eq. (4) generates all the plots using the experimentally measured $\kappa_S = 0.36$, which accounts for the propagation loss in free space after the phase conjugator, collection efficiency of a collimator, and the transmissivities of two optical filters. In the experiment, $N_B$ ranges from $10^4$ to $10^5$, $G - 1 = 0.257 \times 10^{-3}$ so Eq. (6) is fully justified. As such, the exact and the asymptotic results for the BER agree very well.

## III. THEORETICAL MODEL FOR CLASSICAL COMMUNICATION

Here, we present a theoretical model for classical communication (CCOMM) based on the result of Ref. [9]. The output of the ASE source is split into two arms to generate pairs of two-mode Gaussian state across $\hat{a}_S^{(k)}$ and $\hat{a}_R^{(k)}$, where $k \in \{1, M\}$ is the index for the mode pair. After $\theta$-phase encoding by Alice, the covariance matrix of the two-mode Gaussian state becomes

$$V = \begin{pmatrix} 1+N_S & 0 & \sqrt{N_S N_R}e^{i\theta} & 0 \\ 0 & N_S & 0 & \sqrt{N_S N_R}e^{-i\theta} \\ \sqrt{N_S N_R}e^{-i\theta} & 0 & 1+N_R & 0 \\ 0 & \sqrt{N_S N_R}e^{i\theta} & 0 & 1+N_R \end{pmatrix}. \tag{8}$$

The noisy channel transforms the signal mode into $\hat{a}_S'^{(k)} = \sqrt{\kappa}\hat{a}_S^{(k)} + \sqrt{1-\kappa}\hat{a}_B^{(k)}$, and the reference mode suffers a slight experimental loss $\hat{a}_R'^{(k)} = \sqrt{\eta_R}\hat{a}_R^{(k)} + \sqrt{1-\eta_R}\hat{a}_v^{(k)}$, where $\hat{a}_B^{(k)}$ is the environmental background thermal mode that injects $(1-\kappa)\langle \hat{a}_B^{\dagger(k)}\hat{a}_B^{(k)}\rangle = N_B$ number of noise photons, and $\hat{a}_v^{(k)}$ is a vacuum mode. The balanced receiver recombines the two modes and produces at its two output ports modes $\hat{a}_X^{(k)} = (\hat{a}_S'^{(k)} + \hat{a}_R'^{(k)})/\sqrt{2}$, $\hat{a}_Y^{(k)} = (\hat{a}_R'^{(k)} - \hat{a}_S'^{(k)})/\sqrt{2}$. Bob detects the photon number difference summed over *all* $M$ modes, yielding $N \equiv N_X - N_Y$ with $N_X, N_Y$ the random outputs of the measurements on $\sum_{k=1}^{M}(\hat{a}_X^{\dagger(k)}\hat{a}_X^{(k)})$, $\sum_{k=1}^{M}(\hat{a}_Y^{\dagger(k)}\hat{a}_Y^{(k)})$. $N$ is a Gaussian random variable with mean $\mu(\theta) = 2M\sqrt{\eta_R\kappa N_R N_S}\cos(\theta)$ and variance $\sigma^2(\theta) = M[N_B + \eta_R N_R + 2\eta_R N_B N_R + \kappa N_S + 2\eta_R \kappa N_R N_S + 2\eta_R \kappa N_R N_S \cos(2\theta)]$. Hence, the BER for CCOMM with binary phase-shift keying encoding on all $M$ i.i.d. modes is derived as

$$P_e = \frac{1}{2}\text{erfc}(\sqrt{MR_{\text{ASE}}}) = \frac{1}{2}\text{erfc}\left(\sqrt{\frac{M[\mu(0)-\mu(\pi)]^2}{2[\sigma(0)+\sigma(\pi)]^2}}\right). \tag{9}$$

In our experiment, $N_R = 1.2 \times 10^3 \gg 1$, $N_B \gg N_S$, so

$$R_{\text{ASE}} \simeq \frac{\kappa N_S}{N_B + 1/2}. \tag{10}$$

This coincides with the error exponent of homodyne discrimination of binary coherent states. We see that the loss $\eta_R$ on the reference is insignificant for the BER. Note that as the variance of the output signal is much greater than the shot noise, the detector efficiency $\eta_D$ does not change the BER to the leading order. In the experiment, additional loss is introduced at the receiver to avoid saturating the detector. Other experimental nonidealities would however increase the BER. For example, imperfect dispersion matching between the signal and the reference would reduce the interference efficiency of the homodyne receiver. Also, slight deviation from perfect balancing of the BS would lead to additional noise in the homodyne measurements. To account for these experimental imperfections, we introduce a fitting parameter $\kappa_F$ in the error exponent of CCOMM and obtain the following BER formula for CCOMM:

$$P_e = \frac{1}{2}\text{erfc}\left(\sqrt{\frac{M\kappa\kappa_F N_S}{N_B + 1/2}}\right), \tag{11}$$

which is presented as Eq. (4) in the main text.

---




\end{document}